\documentclass[twocolumn,preprintnumbers,amsmath,amssymb]{revtex4}
	\usepackage{graphicx}
	\usepackage{dcolumn}
	\usepackage{bm}

	\usepackage{float,color}
	\usepackage{amsmath}

	\usepackage[colorlinks,linkcolor=blue,citecolor=blue,urlcolor=blue]{hyperref}

\begin{document}

	\title{Aging, computation, and the evolution of neural regeneration processes}

	\author{Aina Oll\'e-Vila$^{1,2}$, Lu\'is F Seoane$^{3}$, Ricard Sol\'e$^{1,2,4}\footnote{ricard.sole@upf.edu}$} 
		\affiliation{$^{1}$ICREA-Complex Systems  Lab, Universitat Pompeu Fabra,   08003   Barcelona}
		\affiliation{$^{2}$Institut de Biologia Evolutiva (CSIC-UPF), Psg Maritim Barceloneta, 37, 08003 Barcelona}
		\affiliation{$^{3}$Instituto de F\'isica Interdisciplinar y Sistemas Complejos IFISC (CSIC-UIB), Campus UIB, 07122, Palma de Mallorca, Spain}	
		\affiliation{$^{4}$Santa Fe Institute, 399 Hyde Park Road, Santa Fe NM 87501, USA}

	\begin{abstract}
		
		Metazoans are capable of gathering information from their environments and respond in predictable ways. These
		computational tasks are achieved by means of more or less complex networks of neurons. Task performance must be
		reliable over an individual's lifetime and must deal robustly with the finite lifespan of cells or with connection
		failure -- rendering aging a relevant feature in this context. How do computations degrade over an organism's
		lifespan? How reliable can computations remain throughout? In order to answer these questions, here we approach the
		problem under a multiobjective (Pareto) optimization approach. We consider a population of digital organisms equipped
		with a neural network that must solve a given computational task reliably. We demand that they remain functional (as
		reliable as possible) for an extended lifespan. Neural connections are costly (as an associated metabolism in living
		beings) and degrade over time. They can also be regenerated at some expense. We investigate the simultaneous
		minimization of the metabolic burden (due to connections and regeneration costs) and the computational error and the
		tradeoffs emerging thereof. We show that Pareto optimal designs display a broad range of potential solutions: from
		small networks with high regeneration rate, to larger and redundant circuits that regenerate slowly. The organism's
		lifespan and the external damage rates are found to act as an evolutionary pressure that improve the exploration of
		the space of solutions and poses tighter optimality conditions.  We also find that large damage rates to the circuits
		can constrain the space of the possible and pose organisms to commit to unique strategies for neural systems
		maintenance. 
	
	\end{abstract}

	\maketitle

	\section{Introduction}
		\label{sec:1}

		A major revolution in the history of multicellular life was the emergence of neurons, a new class of cells that
		enhanced the processing and storage of information beyond the genetic level \cite{jablonka2006evolution}. Such
		revolution enabled fast adaptation to environmental fluctuations. Combining these apt building blocks, a more
		short-sighted sensor-actuator logic was soon backed by more complex networks, resulting in yet further increased
		capabilities for information processing and representation of the external environment \cite{bickerton2018language}.
		Alongside, organism size and longevity also increased. Both required regeneration mechanisms that would sustain
		organismal coherence over extended periods of time, far beyond the cellular lifespan. An alternative (or
		complementary) way to guarantee reliable computation is through an appropriate connectivity pattern -- e.g. faulty
		functioning due to loss of connections could be counterbalanced by redundant wires, among other mechanisms. How this
		can be achieved was early explored by the first generation of mathematicians dealing with unreliable computers
		\cite{von1956probabilistic, winograd1963reliable}. More recent works have addressed some of these questions from the
		perspective of reliability theory \cite{barlow1965mechanism}, particularly in relation to the role played by parallel
		versus sequential topologies \cite{gavrilov2001reliability}.

		The ability to regenerate the nervous system presents an extraordinary variation in Metazoa. Regarding vertebrate
		species, all of them can produce new neurons postnatally in specific regions of their nervous system, but only some
		lower vertebrates (fish and amphibians) can significantly repair several neural structures. Some regenerative ability,
		however, is found also in reptiles and birds, and even in mammals \cite{ferretti2011there}. Remarkably, replacement of
		all or part of the nervous system has been documented in a few invertebrate phyla, including coelenterates, flatworms,
		annelids, gastropods and tunicates \cite{moffet2012nervous}.  Such re-grown neural networks are parsimoniously
		integrated within the rest of the circuitry, stressing how phenotypic functionality is recovered. Although this
		ability is largely absent in higher vertebrates, evidence piles up that the potential might lay dormant
		\cite{levin2007large, levin2009bioelectric, tseng2010induction}. Another open question concerns whether new neurons
		are being created throughout our lives in the absence of damage. While few or no new neurons seem to grow in most
		parts of human brains, there is evidence of limited neurogenesis in the hippocampus \cite{eriksson1998neurogenesis,
		van2002functional} and the olfactory bulb \cite{hack2005neuronal}. Several species, notably fish, present large rates
		of neurogenesis, often requiring apoptosis of older cells to make place for the new ones
		\cite{zupanc2006neurogenesis}. 

		The origin of this diversity of strategies around nervous system maintenance is a major challenge
		\cite{tanaka2009considering}. What are the evolutionary drivers behind each solution? The metabolic cost of wires
		readily comes to mind, and has already been explored as a major constraint of neural architecture
		\cite{chen2006wiring, chklovskii2002wiring, raj2011wiring}, while the regeneration costs are also obvious.  The
		relevance of these factors must be analyzed under the light of a phenotypic function. This, in neural circuits, traces
		back to computations that must be implemented within reasonable error bounds. This computational performance
		constitutes a third dimension relevant to our research questions.

		What is the optimal tradeoff between these factors for reliable neural circuits? Is the range of solutions
		parsimonious, or are there more locally stable designs that hinder the access to other possibilities? In order to
		answer these questions, we asses the maintenance of reliable computations over extended lifespans while enduring an
		aging process (inflicted through an external damage). Interestingly, the interplay between the lifespan of the whole
		organism versus the time scale of its constituent parts brings in an extra factor in our study. Its relevance becomes
		apparent in the empirical record, notably in the apoptosis of older neurons sought by some fish species
		\cite{zupanc2006neurogenesis} -- implying that a valid regeneration strategy actively shortens the useful life of the
		organism's building blocks. How are design spaces affected by these different factors -- computational performance,
		organismal lifespans (and its relation to component time scales), external damages, and the severity of metabolic
		costs? Early and current research has studied how given computational functions are implemented by evolved neural
		networks \cite{miller1989designing, angeline1994evolutionary, yao1999evolving, floreano2008neuroevolution,
		stanley2002evolving, stanley2009hypercube}, but these studies seldom connect wiring costs with possible repair
		processes.

		As noted above, our research questions can only be addressed given a computational task that the underlying circuitry
		must solve. In living organisms this further results in diverse anatomical patterns and neural plans, with the
		computational tasks ingrained in each organism's phenotype. This additional diversity falls beyond the scope of this
		paper. To gain some insight in the tradeoffs behind neural circuits maintenance, we resort to minimal toy models based on
		networks of Boolean units that solve archetypal tasks. Simple Boolean models have been used to explore the basic
		principles of neural functions and the role played by architecture, including information propagation thresholds
		\cite{eckmann2007physics}, locomotion and gating \cite{szekely1965logical, stent1978neuronal, friesen1978neural},
		pattern formation \cite{amari1977dynamics}, or the emergence of modularity \cite{clune2013evolutionary}. In a more
		general context of evolved circuits for artificial agents, evolved neural networks play a central role  in the
		development of biologically inspired robotic systems \cite{floreano2001co, pfeifer2007self}.

		In this spirit, we test feed-forward networks of Boolean units. We set a fixed number of layers, a varying number of
		units in each layer and connections across layers, and a range of regeneration rates. These toy neural circuits are
		tasked with solving a series of computations while responding to the three evolutionary forces outlined above: i) a
		cost stemming from computational errors, ii) another cost associated to wiring, and iii) the cost associated to the
		regeneration of damaged structures. To integrate these evolutionary forces without introducing unjustified biases that
		would assign more importance to a factor over the others, we adopt a (Pareto) Multi-Objective Optimization (MOO)
		approach \cite{coello2006twenty, schuster2012optimization, seoane2016multiobjetive}. This framework explores designs
		that simultaneously minimize all costs involved. The solution of a MOO problem is a restricted region in the space of
		possible designs. This solution embeds the diversity of somehow optimal strategies in a mathematical object whose
		geometry has been linked to phase transitions \cite{seoane2013multiobjective, seoane2015phase, seoane2015systems,
		seoane2016multiobjective, seoane2018morphospace} and criticality \cite{seoane2015systems, seoane2018morphospace}.
		These phenomena give us some insights about how accessible the range of optimal solutions are: whether evolutionary
		biases can navigate them smoothly as they vary, or whether locally optimal designs dominate under discrete value
		regions of the different costs so that changes only happen abruptly. 

		In section \ref{sec:2} we introduce the elements of our toy model: i) the computational tasks explored, ii) the
		implementation of the feed-forward networks and their aging process, and iii) how all of this comes together under
		MOO. In section \ref{sec:3} we go over the results, including how each evolutionary pressure affects the shape of MOO
		solutions in design space, and how this relates to the biology of the problem. Section \ref{sec:4} concludes
		discussing our results within existing literature.

	\section{Methods}
		\label{sec:2}

		\subsection{Computational tasks}
			\label{sec:2.1}
		
			\begin{figure*}[t]
				\includegraphics[width=0.7 \textwidth]{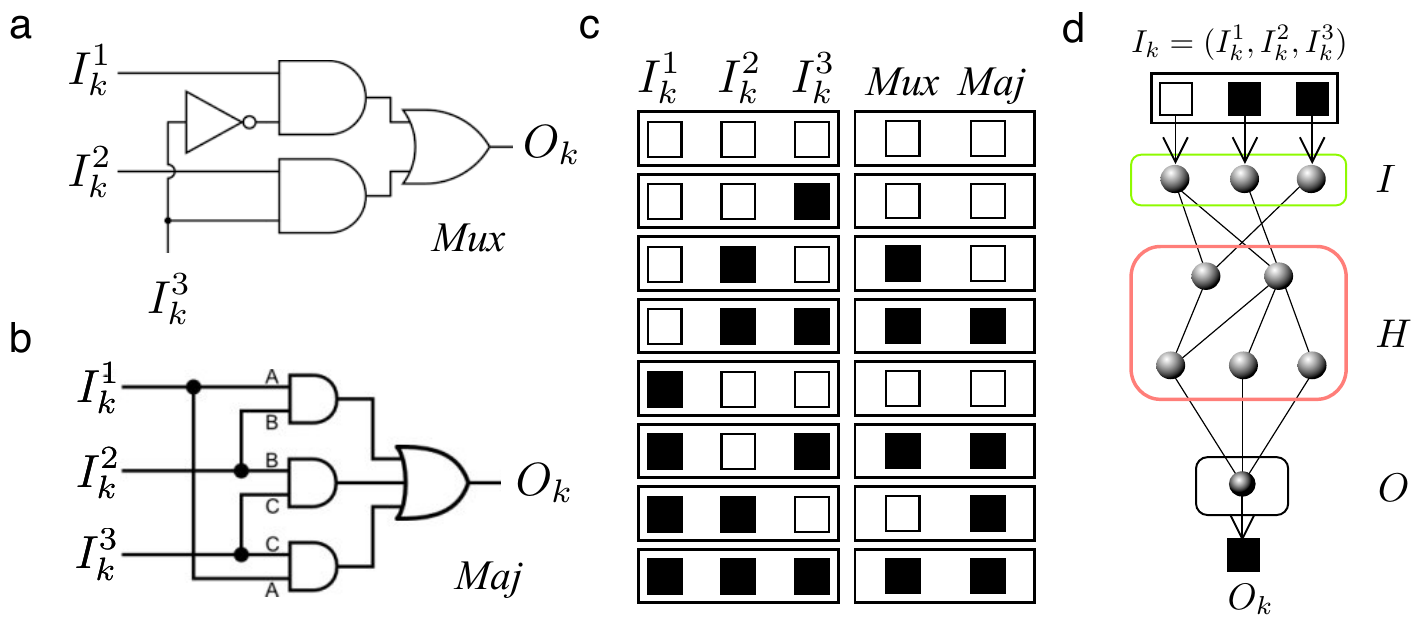}

        \caption{\textbf{Computational tasks and neural implementation}. The agents used in our evolutionary algorithm
are tested with two different, three-input, one output logic functions: (a) the multiplexer ({\em Mux}) and (b) the
majority rule ({\em Maj}). These are traditionally implemented using logic gates as shown. Each circuit has an
associated logic table that fully defines each performed computation. (c) Left, the eight possible binary sets of inputs
$\blacksquare = 0$ and $\square = 1$. Each possible input string, such as $\blacksquare \blacksquare \square$, results
in a unique output. The outputs for {\em Mux} and {\em Maj} are shown (right). (d) In our evolving system, Boolean gates
are replaced by feed-forward neural networks with several layers including input (I), output (O), and hidden (H) ones.}

				\label{fig:figBFunctions}
			\end{figure*}

			We ask our toy neural networks to compute some arbitrary Boolean function: 
				\begin{equation}
					\varphi: \Sigma^N \longrightarrow \Sigma^M, 
					\label{eq:task}
				\end{equation}
			with $\Sigma = \{ 0,1 \}$. In this paper we used $N=3$, $M=1$, and two archetypal computations (figure
			\ref{fig:figBFunctions}): i) The {\em multiplexer} ({\em Mux}) where the ``selector'' bit $i_3$ chooses one of the
			other inputs ($i_1$ or $i_2$) as output. ii) The {\em majority rule} ({\em Maj}), which returns $1$ if most input
			bits are $1$ and $0$ otherwise. {\em Maj} is well known within the study of circuit redundancy and error correction.

		\subsection{Feed-forward neural networks}
			\label{sec:2.2}

			We subjected a population of Boolean, feed-forward neural networks to an evolutionary process that optimizes
			diverse design aspects according to the MOO logic detailed below. Here we clarify the general network
			architecture, how they compute, and some of the diversity allowed.

			Each network consisted of three input units ($I \equiv \{i_1, i_2, i_3\}$), two hidden layers with a varying
			number of units in each layer (from $1$ to $h_{max} = 15$) and connections across layers, and one output unit ($O
			\equiv \{o_1\}$). Each connection carried a weight $\omega_{ij} = \pm 1$, and each unit had an activation
			threshold $\theta_i \in [0.01, 1]$. Networks computed following a McCulloch-Pitts function \cite{rojas2013neural}:
				\begin{equation}
					S_i= H \left( -\theta_i + \textstyle{\sum_{j}} \omega_{ij} S_{j} \right), 
				\end{equation}
			where $j$ runs over input signals to unit $i$, and $H(\cdot)$ represents the Heaviside step function. 

			Thus each network computes a Boolean function. With the variety allowed across networks (different number of
			units, connections, and $\theta_i$), there are several ways to implement a desired function. Different network
			designs will incur in different costs due to the expense of wiring or regeneration (see below), and their
			computational reliability will degrade differently for distinct topologies as they {\em age} (also, see below).
			This constitutes the basis of our MOO exploration.

 		\subsection{Aging process}
			\label{sec:2.3}

			The networks are evaluated over an extended period $t = 1, \dots, \tau$ during which their connections are eroded.
			$\tau$ defines the required durability for the whole network, which would correspond to the lifespan of a modeled
			organism. At the beginning of each evaluation step, each connection is removed with a probability $\delta$, defining
			a {\em damage rate}. Also, the network restores each missing link with probability $\rho$, thus defining a {\em
			regeneration rate}. Recovered connections display their original weight (i.e. some detailed memory is never lost).

			After knocking off some connections and regenerating others, we assess the reliability of each network in computing
			$\varphi$ (eq. \ref{eq:task}, i.e. {\em Mux} or {\em Maj}). An approximate mean-field model (see Supplementary
			Material, section II) shows how the damage/regeneration process eventually results in an average steady number of
			missing connections.

			We performed experiments with different $\delta$ and $\tau$. Notice that $\delta$ defines a damage rate, but also
			contributes to setting an average lifespan ($\tau_{link} \sim 1/\delta$) for the network connections. This, together
			with the timescale of the network lifespan can be combined into a dimensionless ratio $r_\tau \equiv \delta \cdot
			\tau \sim \tau/\tau_{link}$. This ratio is an informative index in analyzing some results, but note that the actual
			$\tau_{link}$ is also affected by regeneration values. 

			This aging process might result in plainly unfeasible networks -- i.e. graphs that become disconnected such that
			information cannot flow from input to output. We track the proportion of time that a network $\gamma$ is thus broken
			through a {\em feasibility function} $F_f(\gamma) \in [0, 1]$:
				\begin{eqnarray}
					F_f(\gamma) &=& {1 \over \tau} \sum_t \xi(\gamma, t); 
				\end{eqnarray}
			where $\xi(\gamma, t) = 1$ if $\gamma$ remains feasible (connections exist from input to output, and from all input
			units to the first hidden layer after damage and regeneration at $t$). Otherwise, $\xi(\gamma, t) = 0$ (check section
			III-E in Supplementary Material for more details).

				\begin{figure*}[ht!]
	
				\includegraphics[width=0.6 \textwidth]{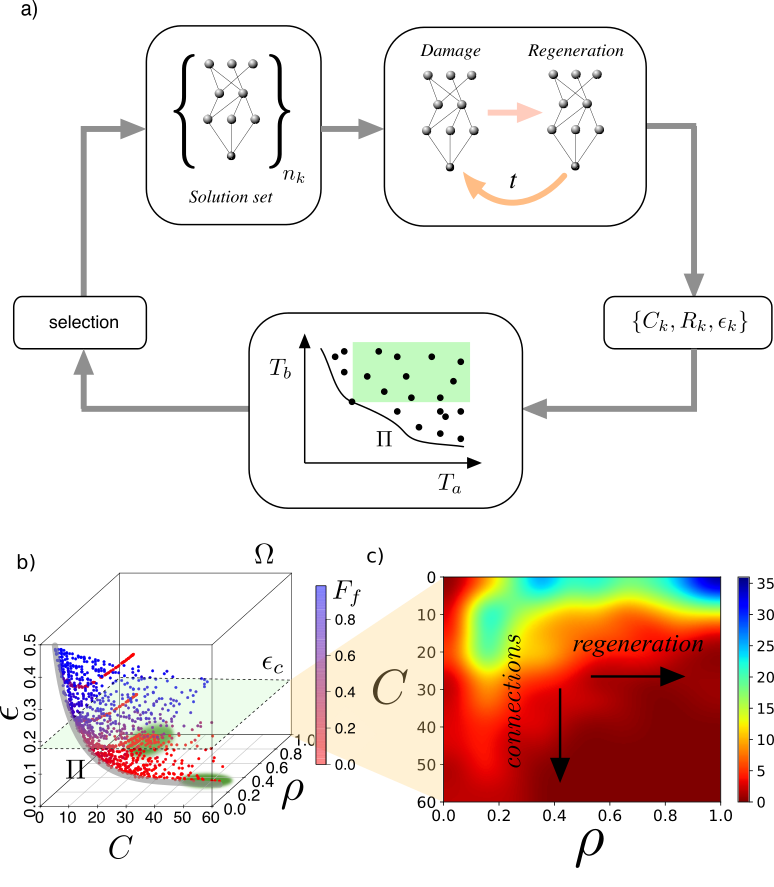}

				\caption{\textbf{Evolutionary algorithm for neural regeneration and output processing}. (a) Iterative process of the
MOO algorithm. A population $\hat{\Gamma}$ of $N=480$ networks $\gamma \in \hat{\Gamma}$ is subjected to a damage: at
each of $\tau$ time steps, their connections are lost with probability $\delta$; meanwhile, the network attempts to
compute. Computational performance of this damaged network ($T_1 \equiv \epsilon(\gamma)$) is measured, along with costs
associated to wiring ($T_2 \equiv C(\gamma)$) and regeneration of lost connections ($T_3 \equiv \rho(\gamma)$). These
{\em optimization targets} should be minimized. Bad solutions have a worse performance in all those targets
simultaneously (green shaded area): they are {\em dominated}. Better network designs cannot perform optimally in all
senses. They are rather better in trading good performance in a target for worse performance in others, thus avoiding
dominance. Less dominated networks at a given iteration of the MOO algorithm are carried to the next generation and used
to produce offspring. This algorithm proceeds for $g^{max}$ iterations. (b) We execute this algorithm ten times for each
fixed conditions of damage ($\delta$) and length of the aging process ($\tau$). This results in a combined population
$\hat{\Gamma}$ whose non-dominated subset $\hat{\Pi}^{end} \subset \hat{\Gamma}$ approximates the optimal tradeoff
between the targets involved. Blurred green circles indicate extreme phenotypes (investment is maximal in regeneration
and minimal in connectivity, or vice versa). A steep relation between the computation error ($T_1$) and the structural
feasibility of the network (as captured by $F_f(\gamma)$, see SM section VII, figures S18-S25) suggests a robust
threshold delimiting acceptable computation $\epsilon_c = 0.2$. (c) Density of network designs in $\hat{\Gamma}^{end}$
over the $\rho-C$ plane that compute acceptably ($\epsilon(\gamma) < \epsilon_c$). These plots will be used to explore
the tradeoff between regeneration and high connectivity. }

				\label{figalgorithm}
			\end{figure*}

		\subsection{Multiobjective Optimization}
			\label{sec:2.4}

			Given the set $\Gamma$ of all allowed networks ($\gamma \in \Gamma$), we seek the subset $\Pi \subset \Gamma$ of
			designs $\gamma^{\pi} \in \Pi$ that simultaneously minimize all relevant costs without introducing artificial
			biases. This subset ($\Pi$) is the solution of a MOO or Pareto optimization problem \cite{coello2006twenty,schuster2012optimization,seoane2016multiobjetive}. Pareto optimal networks $\gamma^{\pi} \in \Pi$ are such that you cannot find two of
			them $\gamma^{\pi}_i, \gamma^{\pi}_j \in \Pi$ with one being better than the other with respect to all
			optimization targets. Pareto optimal designs constitute the best tradeoff possible between the studied traits: we
			can only improve one of them by worsening some other.

			We are invested in the minimization of three optimization targets ($\{T_1, T_2, T_3\}$): 
				\begin{itemize}

            \item[{\bf i}] {\bf Error} $\epsilon(\gamma) \in [0,1]$ in implementing the computational task $\varphi$
(eq. \ref{eq:task}). At each evaluation step ($t = 1, \dots, \tau$), after damage and regeneration, the network
$\gamma$ is fed all $2^M$ possible input bits $I_i$ ($i=1, \dots, 2^3$ for {\em Mux} and {\em Maj}, figure
\ref{fig:figBFunctions}). The network output $O^\gamma(I_i, t)$ is then compared to $\varphi(I_i)$ to compute the
average number, over inputs and network lifespan, of mistaken outputs:
							\begin{eqnarray} 
								T_1 \equiv \epsilon(\gamma) &=&  {\sum_t \sum_{i=1}^{2^M} 
														\left|\left| \varphi(I_i) - O^\gamma(I_i, t) \right|\right| \over \tau \cdot 2^M}, 
							\end{eqnarray}  
where $||\cdot||$ represents absolute value. We average $T_1$ over $10$ independent realizations of the aging process. 

            \item[{\bf ii}] {\bf Connectivity} given by the number of links at $t=0$ before any damage:
							\begin{eqnarray}
								T_2 \equiv C(\gamma) &=& \sum_{\omega} \left(1 - \delta_{\omega, 0}\right), 
							\end{eqnarray} 
where $\delta_{\omega, 0}$ is Kronecker's delta and the sum runs over all possible weights such that each non-zero
connection incurs in one unit cost. This target embodies a metabolic burden entailed by wiring costs.

            \item[{\bf iii}] Each network is created with a unique {\bf regeneration rate}:
							\begin{eqnarray}
								T_3 \equiv \rho(\gamma) &\in& [\rho^{min}, 1]. 
							\end{eqnarray}
$\rho(\gamma)$ specifies the probability with which the network recovers each damaged link at each $t=1, \dots, \tau$ of
the aging process. A lower bound $\rho^{min}= 0.01 > 0$ was chosen as such minimum regeneration power seems to be always
present in living systems \cite{tanaka2009considering}.

				\end{itemize}

			The set $\Gamma$ of possible networks is combinatorially vast. To explore it, we resort to an evolutionary MOO
			algorithm (figure \ref{figalgorithm}). For a seed (random) population $\hat{\Gamma}(0) \subset \Gamma$ of networks
			$\gamma \in \hat{\Gamma}(0)$ we evaluate their performance in each target $\{T_1(\gamma), T_2(\gamma),
			T_3(\gamma)\}$, select the fittest ones according to a Pareto dominance criterion, and produce diverse offspring
			through crossover and mutation. This evolves our network ensemble towards the MOO solution $\Pi$ over generations
			$g = 0, \dots, g^{max}$.

			Our {\em target space} (with $\{T_1, T_2, T_3\}$ as axes, figure \ref{figalgorithm}{\bf b}) allows us to compare
			networks without favoring performance in any $T_k$ over the others: Given $\gamma_i, \gamma_j \in \Gamma$, we say
			that $\gamma_i$ {\em dominates} $\gamma_j$ (and note it $\gamma_i \prec \gamma_j$) if $\gamma_i$ is no worse than
			$\gamma_j$ in all targets ($T_k(\gamma_i) \le T_k(\gamma_j) \forall k=1, 2, 3$) and it is strictly better in at least
			one target ($\exists k', T_{k'}(\gamma_i) < T_{k'}(\gamma_j)$).

			Using these guidelines (following \cite{konak2006multi}), we conducted an MOO with a population of $N=480$ networks
			over $g^{max} = 4000$ generations. At each generation, $\{T_1(\gamma), T_2(\gamma), T_3(\gamma)\}$ were used to
			calculate dominance scores: $D(\gamma_i, g) \equiv ||\{\gamma_j \in \hat{\Gamma}(g), \gamma_j \prec \gamma_i\}||$ --
			i.e. the number of designs $\gamma_j \in \hat{\Gamma}(g)$ that dominate $\gamma_i \in \hat{\Gamma}(g)$. Pareto
			optimal networks have $D(\gamma^{\pi} \in \Pi) = 0$ (but not the other way around). Poor-performing networks soon
			become dominated by others. We {\em copied} all $\gamma_i$ with $D(\gamma_i, g)=0$ into $\hat{\Gamma}(g+1)$. This
			{\em elitist} strategy ensures that we never lose the fittest designs. The half of the population with largest
			$D(\gamma_i, g)$ is discarded. Networks in the remaining half are combined to bring $\hat{\Gamma}(g+1)$ to its full
			size ($N=480$). The resulting children are mutated (see Supplementary Material, section III, for further
			implementation details). We ran this algorithm $10$ times for each $(\delta, \tau)$ condition. The final population
			of all $10$ runs are combined in a unique set noted $\hat{\Gamma}^{end} \equiv \hat{\Gamma}(g = g^{max})$. The
			results reported correspond to the Pareto-optimal networks of this merged data set, $\hat{\Pi}^{end} \equiv
			\{\gamma_k \in\hat{\Gamma}^{end}, D(\gamma_k) = 0\}$ (see Supplementary Material, section IV, to observe all the
			final results for each $(\delta, \tau)$ condition). We assume $\hat{\Pi}^{end} \simeq \Pi$, but full convergence
			cannot be guaranteed. 

			Figure \ref{figalgorithm}{\bf b} illustrates this for particular $(\delta, \tau)$ conditions. $\hat{\Pi}^{end}$ (and
			actually $\Pi$ itself) includes designs that compute very badly (large $T_1 \equiv \epsilon(\gamma)$) but have been
			selected because of their negligible regeneration cost and number of links. Pareto dominance offers no principled way
			to dispose of these networks, even though they would fade away in a biological setting because they plainly fail to
			function.  Interestingly, we observed that computational errors are often associated to deeper structural breakdowns,
			measured by $F_f$. Figure \ref{figalgorithm}{\bf b} shows, color-coded, the feasibility $F_f(\gamma)$ of each
			network. Those with large computational errors often cannot even convey information from input to output. Plotting
			$\epsilon(\gamma)$ vs $F_f(\gamma)$  (see Supplementary Material, section VII) we noted that this degradation of
			computation capabilities and  network structure follows a logistic curve with a marked threshold $\epsilon_c \sim
			0.24$. This value turned out to be similar across realizations of the algorithm and for different $(\delta, \tau)$
			conditions (both for \emph{Mux} and \emph{Maj} functions; see Supplementary Material, figures S18 to S25). We took it
			as a heuristic limit (horizontal plane in figure \ref{figalgorithm}{\bf b}) to select {\em acceptably working}
			designs (we took $\epsilon_c$ = 0.2 to be on the safe side). Figure \ref{figalgorithm}{\bf d} shows all networks with
			a performance better than this threshold ($T_1 \equiv \epsilon(\gamma) < \epsilon_c = 0.2$) in a $C - \rho$ map. (See
			Supplementary Material, section V, to observe the density plots for each $(\delta, \tau)$ condition, section I, to check on 	all the parameters of the model, and section III to check on further
			implementation details.)

	\section{Results}
		\label{sec:3}

		\subsection{Tradeoff between connectivity and regeneration}
			\label{sec:3.1}

			\begin{figure*}[t]
				\includegraphics[width=0.8 \textwidth]{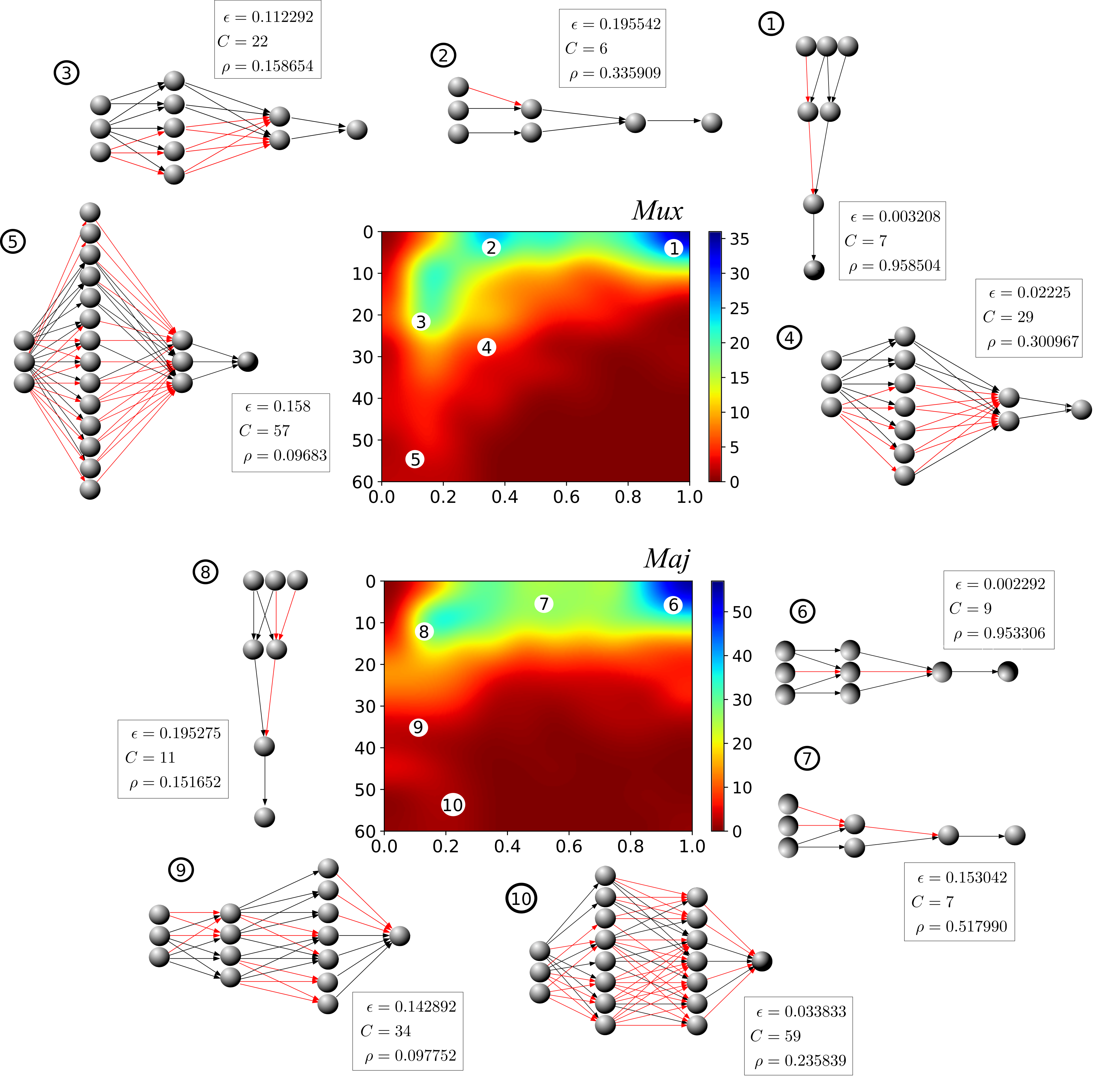} 
		
        \caption{\textbf{Connectivity-regeneration ($C-\rho$) tradeoff as a means to achieve reliable computation.} Two
selected density plots for {\em Mux} (top) and {\em Maj} (bottom) of solutions $\gamma \in \hat{\Gamma}^{end}$ with
reliable computation ($\epsilon(\gamma) < \epsilon_c$) throughout the $\rho - C$ plane. Alongside, representative
networks from characteristic regions of phenotype space. Performance of these networks in all three targets is shown
(enclosed squares). Results obtained for $\delta = 0.04$, $\tau = 300$ for {\em Mux} and $\delta = 0.02$ and $\tau =
1500$ for {\em Maj}. (See SM for other values.)}

				\label{fig1}
			\end{figure*}

			Figure \ref{fig1} shows abundance of networks $\gamma \in \hat{\Pi}^{end}$ with $\epsilon(\gamma) < \epsilon_c \sim
			0.2$ (i.e. Pareto optimal designs that further satisfy the reasonable phenotypic performance $\epsilon_c$, which also
			entails a persisting structural integrity -- large feasibility $F_f(\gamma)$) for a low damage $\delta$ condition,
			both for \emph{Mux} and \emph{Maj} as computational tasks. The plotted abundance of designs across the $\rho - C$
			plane captures the tradeoff between connectivity ($T_2 \equiv C(\gamma)$) and regeneration ($T_3 \equiv
			\rho(\gamma)$). A broad range of optimal solutions compatible with proper functionality is showcased. 

			Both panels (and further plots in SM) show a higher density of solutions around areas with less connections and higher
			regeneration (e.g. peak in the upper right corner, figure \ref{fig1}{\bf a}). Graphs labeled $1$ and $2$ for {\em
			Mux} (figure \ref{fig1}{\bf a}) and $6$, $7$, and $8$ for {\em Maj} (figure \ref{fig1}{\bf b}) illustrate minimal
			circuits implementing those tasks. The density of designs found with lower regeneration rates (which demands higher
			connectivity) is notably smaller in comparison. Some graphs (3, 4, and 5 for {\em Mux}; 9 and 10 for {\em Maj}) sample this more sparsely occupied region with lower
			regeneration levels and more densely connected circuits. 

			Trends are shared among the two Boolean functions tested, such as the lower density of solutions in this region of
			phenotype space, and the resilience that is obtained trough abundant, duplicated links. This hints us at general
			patterns emerging despite the potential variety in phenotypes imposed by different computational tasks. 

		\subsection{Network lifespan and external damage act as evolutionary pressures}
			\label{sec:3.2}
			\begin{figure*}[t]
				\includegraphics[width=0.8 \textwidth]{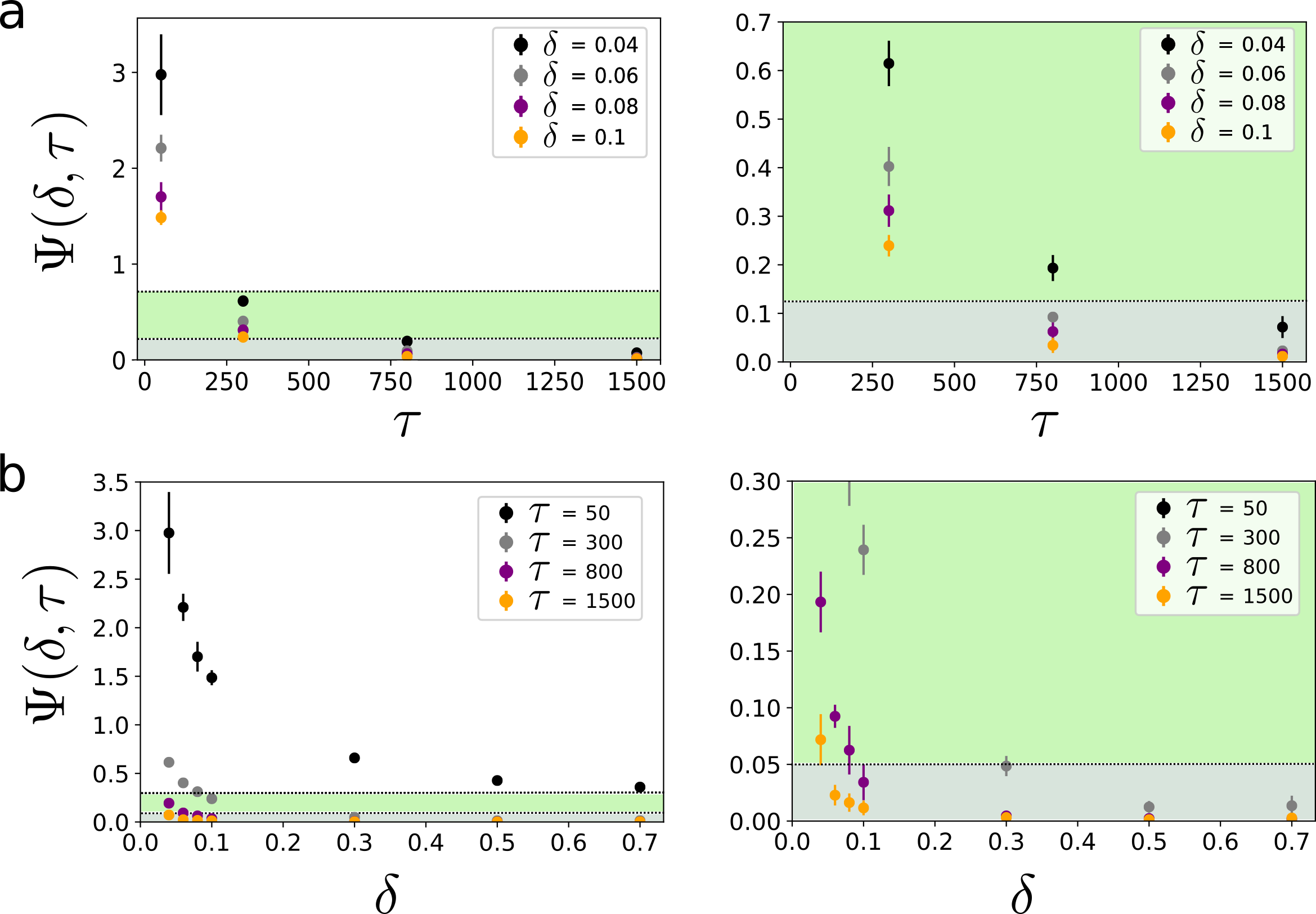} 

        \caption{\textbf{Life time $\tau$ and damage rate $\delta$ act as evolutionary pressures.} (a) Ratio $\Psi
(\delta, \tau)$ between the number of non-Pareto and Pareto-optimal solutions in $\hat{\Gamma}^{end}$ as a function of
the $\tau$ for fixed values of damage rate $\delta$ (for \emph{Mux}). $\Psi$ decreases for increasing $\tau$. Only a
sample of $\delta$ values (0.04, 0.06, 0.08 and 0.1) is shown to help visualization (other conditions, in SM section VI,
also for \emph{Maj}; all results in Supplementary Material support the conclusions in the main text). (b) Same ratio for fixed values of
$\tau$ and varying $\delta$. Again, $\Psi$ drops as $\delta$ increases. Decreases in $\Psi$ capture a higher
evolutionary pressure towards Pareto optimality. }

				\label{fig4}
			\end{figure*}

			As discussed above, our aim is to asses the influence of additional distinct features on our evolutionary framework,
			namely the lifespan of networks and the damage inflicted to their links during the aging process. The former offers a
			window to the effects of life length ($\tau$) on the selection process whereas the latter, implemented by $\delta$,
			allows including the stochastic perturbations that hamper proper computation. What is the impact of lifetime ($\tau$)
			or damage rates ($\delta$) in the optimal space of solutions resulting from our model? 

			We have found that, for a fixed damage rate, longer agent life times act as an evolutionary pressure, such that the
			algorithm better explores the optimal tradeoff by pushing harder our evolving population against it. The same happens
			in the opposite case, e.g. increasing damage rates for fixed life times. To capture this, we compute the amount of
			non-dominated versus dominated solutions in $\hat{\Gamma}^{end}$:
				\begin{eqnarray}
					\Psi (\delta, \tau) &=& {||\{\gamma_k \in \hat{\Gamma}^{end}, D(\gamma_k) > 0\}|| 
																					\over ||\{\gamma_k \in \hat{\Gamma}^{end}, D(\gamma_k) = 0\}||} \nonumber \\ 
															&=& {||\hat{\Gamma}^{end} - \hat{\Pi}^{end} || \over ||\hat{\Pi}^{end}||}, 
				\end{eqnarray}
			where $D(\cdot)$ is again the dominance score. Each of our evolution takes place under fixed $(\delta, \tau)$
			conditions. Those instances that result in larger $\Psi$ retain more network designs that are suboptimal (in the
			Pareto sense) with respect to other surviving designs. This is, such $(\delta, \tau)$ settings are less strict
			during selection, such that networks that perform relatively worse in all targets simultaneously are still retained.
			Opposed to this, $(\delta, \tau)$ values resulting in a smaller $\Psi$ are more severe regarding Pareto optimality
			selection -- i.e. evolutionary pressure to select Pareto optimal solutions is higher. See Fig. \ref{fig4}a to check
			the effect of fixed damage rates and increasing life times, and Fig. \ref{fig4}b to illustrate the effect of fixed
			lifetime and increasing damage. However, both effects are present in both plots. (Check Supplementary Material,
			section VI, to check the rest of parameter values either for \emph{Mux} and \emph{Maj}.)
			
			Which are the reasons for such observation? Despite both network lifespan and damage rate exert a pressure of the
			same nature (as measured by $\Psi (\delta, \tau)$, the ultimate reasons for such observation might be different. An
			explanation for large $\tau$ as an evolutionary pressure for Pareto optimality could lay in the fact that longer life
			times are synonym of dealing with more extended and numerous threats entailing a larger information retrieval from
			the environment. Therefore, improvements during evolutionary time can have a larger impact on this longer living
			populations compared to shorter living ones. Regarding the influence of the degree of damage inflicted to networks
			($\delta$), the reason for such evolutionary pressure is probably rooted to the inherent harsher survival conditions
			imposed in larger damage regimes.
					
			Importantly, the interplay of both features might also be acting as an evolutionary pressure (as measured by $\Psi
			(\delta, \tau)$). As presented in section \ref{sec:2.3}, damage rates and network lifespans can be collapsed into a
			dimensionless ratio, $r_\tau \equiv \delta \cdot \tau \sim \tau/\tau_{link}$, as the lifespan of the connections
			$\tau_{link}$ grows monotonously with $1/\delta$, thus capturing a relationship between time-scales proper of whole
			organisms versus those of its parts. An increase of this ratio also entails lower values of $\Psi (\delta, \tau)$,
			meaning that an increase in the difference between the lifespan of the components versus that of the organism might
			be playing a role in the described observations.

		\subsection{Damage rates influence the overall shape of the optimal tradeoff } 
			\label{sec:3.0}

			\begin{figure*}[t]
				\includegraphics[width=\textwidth]{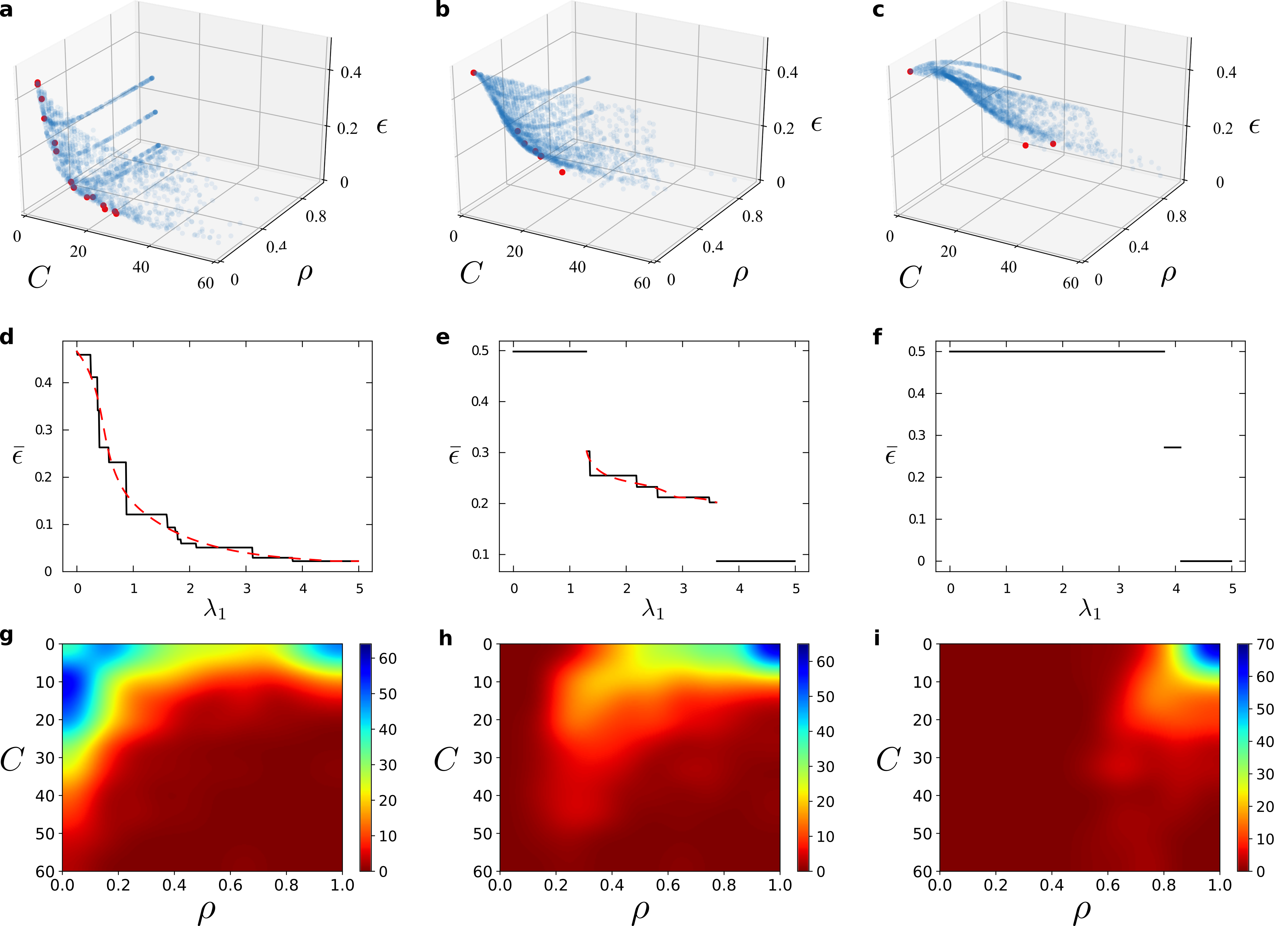} 
		
        \caption{\textbf{Overall shape of the optimal tradeoff and accessibility of phenotype space.} Pareto optimal
strategies $\{\gamma \in \hat{\Pi}^{end}\}$ under fixed conditions ((a) $\delta=0.01$, $\tau=1500$; (b) $\delta=0.1$,
$\tau=300$; (c) $\delta=0.7$, $\tau=50$) plotted in {\em target space}. Each blue dot represents $\{T_1(\gamma),
T_2(\gamma), T_3(\gamma)\}$ for a given $\gamma \in \hat{\Pi}^{end}$. The shape of each embedding surface determines how
accessible phenotype space is if minimizers of a global utility function $\Omega = \sum \lambda_i T_i$ were sought. Red
dots represent such global optima as $\lambda_2=0.009$ and $\lambda_3=2$ are kept fixed and $\lambda_1 \in [0, 5]$.
Increasing damage ((a) $\delta=0.01$, (b) $\delta=0.1$, (c) $\delta=0.7$) results in increasingly more rugged tradeoffs.
Smooth tradeoffs (a) are sampled more evenly by global optimizers, so that successive global optima are similar to each
other. Whatever property we plot of them, e.g. their computational error $\epsilon(\gamma)$ (d, solid black lines),
varies relatively smoothly with $\lambda_1$. Some discreteness remains due to the numerical nature of our experiments
(dashed red lines illustrate the underlying continuous dependency). The corresponding density plot (g) with solutions
fulfilling $\epsilon(\gamma) < \epsilon_c$ shows the tradeoff $\rho-C$ available for the low damage regime (already
shown in figure \ref{fig1}). Tradeoffs with cavities (b, c) leave some phenotypes unsampled by global optima. These
often take leaps in phenotype space, resulting in sudden jumps in any measured property of such solutions as a function
of $\lambda_1$ (e, f). These plots show how a heightened damage rate has an ability to hamper access to existing
phenotypes and fracture the continuity of global optimal designs. The corresponding density plots (h, i) with solutions
fulfilling $\epsilon(\gamma) < \epsilon_c$ show the constraints imposed by increasing damage regimes in the $C-\rho$
projection. Crucially, these available designs can be further constrained by the fracture of the continuity of global
optimal designs. See S2-S7 in SM for further evidence linking ruggedness of the optimal tradeoff to increased damage
rate, both for {\em Mux} and {\em Maj} functions.}

				\label{fig:pFronts}
			\end{figure*}	

			The resulting $\hat{\Pi}^{end}$ embody the optimal tradeoff between all targets involved. This optimal tradeoff can
			be visualized if plotted in target space, where its shape can be different for each $(\delta, \tau)$ fixed
			conditions. Our exhaustive exploration of $(\delta, \tau)$ combinations (see SM section IV) shows a clear overall
			change in the shape of this optimal tradeoff as damage increases -- a change, furthermore, that has consequences in
			terms of phenotype space accessibility and exploration as we discuss below. Figure \ref{fig:pFronts}{\bf a-c} shows
			networks in $\hat{\Pi}^{end}$ for increasing damage regimes. Designs with $\epsilon(\gamma) > \epsilon_c$ are
			retained here. $\hat{\Pi}^{end}$ appears smooth for the low damage regime ($\delta = 0.01$), meaning that the optimal
			tradeoff does not present large cavities or singular points when plotted in target space (figure
			\ref{fig:pFronts}{\bf a}). This surface becomes more rugged (i.e. its curvature changes, thus generating cavities)
			for increasing damage rates ($\delta=0.1, 0.7$, figure \ref{fig:pFronts}{\bf b, c}). 

			This varying ruggedness of optimal tradeoffs can tell us something about how accessible our space of optimal
			solutions is. We could weight all targets linearly into a global optimization function: $\Omega(\gamma, \Lambda) =
			\lambda_1 T_1(\gamma) + \lambda_2 T_2(\gamma) + \lambda_3 T_3(\gamma)$, where $\Lambda \equiv \{\lambda_1, \lambda_2,
			\lambda_3\}$ represent explicit evolutionary biases towards a specific target. For example, a large $\lambda_2$
			versus low $\lambda_{1,3}$ indicates that, in a given environment, minimizing the initial metabolic cost of links is
			outstandingly important. Giving values to the $\lambda_{k}$ could thus define a specific set of external
			environmental pressures. Then the minimization of $\Omega(\gamma, \Lambda)$ selects one single solution
			$\bar{\gamma}(\Lambda) \in \hat{\Pi}^{end}$ out of the Pareto optimal tradeoff. This is, the imposition of such
			specific constraints would force evolution towards a determined region of phenotype space. 
		 
			Cavities and singular points in $\Pi$ ($\sim \hat{\Pi}^{end}$) have been linked to phase transitions
			\cite{seoane2013multiobjective, seoane2015phase, seoane2015systems, seoane2016multiobjective, seoane2018morphospace}
			and critical phenomena \cite{seoane2015systems, seoane2018morphospace} that arise as the $\lambda_k$ are varied --
			i.e. as potential biases in a niche change (perhaps over time, or perhaps because the species has left that niche). A
			phase transition in our problem would indicate that certain network designs are persistently stable under a range of
			external evolutionary conditions, and that a switch from a network design to another would sometimes be drastic even
			if those external conditions would vary just slightly.

			Such cavities and singular points are absent for the low $\delta$ example in figure \ref{fig:pFronts}{\bf a} (the
			same regime was shown in figure \ref{fig1}), meaning that the whole phenotypic space of nervous system maintenance
			strategies can be smoothly visited as evolutionary pressures vary. To illustrate this we have set $\lambda_2$ and
			$\lambda_3$ to a fixed value while varying $\lambda_1 \in [0, 5]$. This means that we sample evolutionary pressures
			that initially disregard good computation but progress towards situations in which computing correctly becomes more
			pressing. Network designs $\bar{\gamma} \in \hat{\Pi}^{end}$ that minimize $\Omega(\gamma, \Lambda)$ for the range of
			$\lambda_1$ are displayed in red in figure \ref{fig:pFronts}{\bf a}. Figure \ref{fig:pFronts}{\bf d} shows the
			computational error for these absolute optima $\bar{\epsilon} \equiv \epsilon(\bar{\gamma}(\Lambda))$	as a function
			of $\lambda_1$. The numerical nature of our experiments introduces some unavoidable discreteness in
			$\bar{\epsilon}(\lambda_1)$; but overall we can see how $\hat{\Pi}^{end}$ is rather smoothly sampled (as compared to
			the next cases) and optimal solutions progress parsimoniously into each other as external evolutionary pressures
			change. 

			Additionally, figure \ref{fig:pFronts}{\bf a} shows how $\hat{\Pi}^{end}$ becomes virtually flat for $T_1 \equiv
			\epsilon \rightarrow 0$. Decrementally small improvements in computation can only be achieved by increasingly larger
			investment on $T_2 \equiv C$ and $T_3 \equiv \rho$. Perfect computation ($\epsilon = 0$) could be reached ultimately.
			But this plot reveals a decreasing return for low errors. Organisms eventually need exaggerated investments to
			achieve negligible computational improvements. This is reflected by a sparse sampling of those costly areas of
			phenotype space, also reflected in the plots of abundance of designs across the $\rho - C$ plane (figure \ref{fig1}).

			Optimal tradeoffs present cavities for the examples in larger damage regimes (figures \ref{fig:pFronts}{\bf b, c}).
			Red dots show absolute optima as the same values of $\Lambda$ are sampled. The cavities in $\hat{\Pi}^{end}$ entail
			that a lower number of different absolute optima are recovered for $\delta=0.1$, and even less for $\delta= 0.7$
			under the same procedure. This is so because certain phenotypes remain optimal over a wide range of $\lambda_1$, thus
			preventing us from visiting part of phenotypic space. Furthermore, when such paramount designs stop being global
			optima, the new preferred network is found far apart in $\hat{\Pi}^{end}$. A small change in $\lambda_1$ would then
			demand a prompt yet drastic adaptation to accommodate the new best. This is reflected in $\bar{\epsilon}(\lambda_1)$
			plots (figures \ref{fig:pFronts}{\bf e-f}) through huge improvements in performance as the bias towards minimizing
			the computational error ($\lambda_1$) increases.

			Notice that in these cases (figures \ref{fig:pFronts}\textbf{b, c}) the overall shape of the Pareto optimal front is
			also constraining to a much reduced area the set of designs which retain reliable computations ($\epsilon <
			\epsilon_c$, see the net effect on the $\rho-C$ density plots in figures \ref{fig:pFronts}\textbf{h, i}), if compared
			with the low damage rate regime shown in figures \ref{fig1} and \ref{fig:pFronts}\textbf{a,g}. Thus, increased damage
			(besides constraining  the available phenotypic space due to emerging phase transitions) also results in a much more
			narrower space when looking only at {\em acceptably performing} designs, disregarding of any global optimization of
			$\Omega$. 

			Importantly, it cannot be discarded that either the network lifespan ($\tau$) or the interplay between it and the
			lifespan of the connections ($\tau_{link}$), measured through $r_{\tau}$, may also have an influence on the change in
			the overall shape of optimal tradeoffs ($r_{\tau} = 15, 30, 35$ in figures \ref{fig:pFronts}\textbf{a,b,c},
			respectively). However, the observations retrieved from the set of tested parameters (see Supplementary Material,
			section IV) show that the clear driver of the observed effect is the increase in damage rates regimes.

	\section{Discussion}
		\label{sec:4}

		In this paper we attempt to provide insights on which are the evolutionary pressures or drivers that may underpin the
		evolution of nervous systems and its regeneration capabilities. Due to the generality of the model, we do not aim to
		answer specific questions but rather extract general principles which might pave the way to future research
		directions. The simplifications we make are considerable, starting from the use of artificial feed-forward neural
		networks to represent a nervous system.  Regarding the dynamics of the system, the so-called aging process to which
		our networks are exposed and their subsequent response is a simplification of the processes of axonogenesis and
		neurogenesis \cite{tanaka2009considering} observed in real nervous systems. While only axonogenesis is explicitly
		incorporated in the model, neurogenesis can be considered to be present as a secondary response linked to the recovery
		of connections (whose loss can cause the breakdown of neuron functionality). On the other hand, it is also known that
		regeneration response is dependent on anatomical location and type of damage \cite{tanaka2009considering}, but such a
		feature has not been incorporated in any way, being another simplification of the model. And yet our schematic
		framework retains what we think are some key factors to understand the evolutionary drivers of neural systems
		maintenance. We can thus disentangle some pressures at play as well as isolate relevant factors for future studies.
		This would probably be more difficult with more complicated models.

		First, our results show that, in the domain of reliable computation, a tradeoff might be at play between high density
		of connections and high regeneration capabilities. This tradeoff shapes the phenotypic space of possible designs for
		network maintenance, and it reminds us of the variety of such strategies in the natural world. The region of phenotype
		space with higher connectivity presumably achieves robust computation through specific redundant connections or degenerate mechanisms \cite{tononi1999measures,edelman2001degeneracy} 
		The alternative, which relies on large
		regeneration rates, is found in the region of phenotype space where networks are smaller. We observe that these are
		also the networks preferentially explored by our MOO algorithm. This is so even if, intuitively, a higher number of
		connections could result in a potentially larger amount of different architectures that solve a same task.

		This suggests that regeneration can be a more reliable strategy -- at least at the scales explored. Strategies betting
		on duplicate pathways might need to deal, e.g., with emerging interactions between surviving connections as faulty
		ones are removed. Such problems could be bypassed by alternate computational paradigms -- e.g. distributed computation
		\cite{macia2012distributed, regot2011distributed} or reservoir computing \cite{jaeger2001echo, maass2002real,
		jaeger2007special, verstraeten2007experimental, lukovsevivcius2012reservoir}. The latter can take explicit advantage
		of such emergent behaviors. Such alternatives could unlock further regions of phenotypic space with a high number of
		similar solutions. But such computational strategies depend crucially on huge numbers of units, and might be
		unreachable for our experiments; thus returning us to the observed bias towards regeneration. 
 
		Secondly, we have found that both the extent of external damage and the network lifespan are acting as active
		evolutionary pressures over the whole population so that it eventually contains more Pareto optimal individuals. The
		increase of both factors gives rise to more strict evolutionary scenarios, in which the pressure to select Pareto
		optimal solutions is higher, resulting in more diverse optimal strategies (better exploration of the optimal
		tradeoffs).  Interestingly, the relationship $r_\tau$ between organismal lifetimes ($\tau$) and the timescale of its
		component parts ($\tau_{link} \sim 1/\delta$) is also suggested to be acting as an evolutionary pressure towards this
		direction. Focus on this ratio becomes a very enticing research avenue if we wonder at what level (organismal versus
		component part) can a Darwinian process store the information gathered as evolution proceeds.
 	
 		Third, from the observations retrieved we have found that regimes of increasing damage (which can be considered an
 		ecological factor) result in more rugged tradeoffs. Such tradeoffs present cavities, which is the characteristic
 		signature of first order phase transitions \cite{seoane2013multiobjective, seoane2015phase, seoane2015systems,
 		seoane2016multiobjective, seoane2018morphospace}. Under varying evolutionary biases, the presence of such transitions
 		can result in history dependency effects similar to hysteresis \cite{sole2011phase}. In such phenomena, evolutionary
 		pressures confronted by a species would vary just slightly and yet a preferred global optimum would change
 		drastically. The species might need to adapt swiftly and retain suboptimal aspects due to frozen accidents, thus
 		resulting in increased evolutionary path dependency. 

		All of this suggests that i) discrete phenotypic space, ii) drastic changes expected under varying external
		evolutionary pressures, iii) phenotypic space becoming less accessible, and iv) heightened path dependency in
		evolution should all become more prominent as the external erosion of our system is higher. Overall, this would mean
		that higher damage rates could induce organisms to commit to specific nervous systems maintenance strategies,
		potentially renouncing an ability to switch options with relative ease. Damage could change swiftly as living beings
		migrate to more benign environments or are suddenly locked on harsher conditions. Similarly to an inherent shorter
		life-time of component parts, a heightened external damage rate has the ability to harshen the strictness of the
		selection process (again, in a Pareto optimality sense; thus implying population-wide phenomena) and also of promptly
		shifting the shape of the optimal tradeoff. Future work will be required to explore these results in a more general
		context of multicellularity (natural and synthetic) where cognitive complexity is a well-defined dimension
		\cite{olle2016morphospace}.

	\begin{acknowledgments}

		The authors thank the members of the Complex Systems Lab for useful discussions, specially Jordi Pi\~nero. This work
		has been supported by an ERC Advanced Grant Number 294294 from the EU seventh framework program (SYNCOM), by the Bot\'in Foundation by Banco Santander through its Santander Universities Global Division, a MINECO grant FIS2015-67616 
		fellowship co-funded by FEDER/UE, by the Universities and Research Secretariat of the Ministry of Business and Knowledge of the Generalitat
		de Catalunya and the European Social Fund, and by the Santa Fe Institute.

	\end{acknowledgments}

	\section{References}
		\bibliographystyle{rsc} 
		\bibliography{regeneration} 

\providecommand*{\mcitethebibliography}{\thebibliography}
\csname @ifundefined\endcsname{endmcitethebibliography}
{\let\endmcitethebibliography\endthebibliography}{}
\begin{mcitethebibliography}{54}
\providecommand*{\natexlab}[1]{#1}
\providecommand*{\mciteSetBstSublistMode}[1]{}
\providecommand*{\mciteSetBstMaxWidthForm}[2]{}
\providecommand*{\mciteBstWouldAddEndPuncttrue}
  {\def\EndOfBibitem{\unskip.}}
\providecommand*{\mciteBstWouldAddEndPunctfalse}
  {\let\EndOfBibitem\relax}
\providecommand*{\mciteSetBstMidEndSepPunct}[3]{}
\providecommand*{\mciteSetBstSublistLabelBeginEnd}[3]{}
\providecommand*{\EndOfBibitem}{}
\mciteSetBstSublistMode{f}
\mciteSetBstMaxWidthForm{subitem}
{(\emph{\alph{mcitesubitemcount}})}
\mciteSetBstSublistLabelBeginEnd{\mcitemaxwidthsubitemform\space}
{\relax}{\relax}

\bibitem[Jablonka and Lamb(2006)]{jablonka2006evolution}
E.~Jablonka and M.~J. Lamb, \emph{Journal of Theoretical Biology}, 2006,
  \textbf{239}, 236--246\relax
\mciteBstWouldAddEndPuncttrue
\mciteSetBstMidEndSepPunct{\mcitedefaultmidpunct}
{\mcitedefaultendpunct}{\mcitedefaultseppunct}\relax
\EndOfBibitem
\bibitem[Bickerton(2018)]{bickerton2018language}
D.~Bickerton, \emph{Language and species}, University of Chicago Press,
  2018\relax
\mciteBstWouldAddEndPuncttrue
\mciteSetBstMidEndSepPunct{\mcitedefaultmidpunct}
{\mcitedefaultendpunct}{\mcitedefaultseppunct}\relax
\EndOfBibitem
\bibitem[Von~Neumann(1956)]{von1956probabilistic}
J.~Von~Neumann, \emph{Automata studies}, 1956, \textbf{34}, 43--98\relax
\mciteBstWouldAddEndPuncttrue
\mciteSetBstMidEndSepPunct{\mcitedefaultmidpunct}
{\mcitedefaultendpunct}{\mcitedefaultseppunct}\relax
\EndOfBibitem
\bibitem[Winograd and Cowan(1963)]{winograd1963reliable}
S.~Winograd and J.~D. Cowan, \emph{Reliable computation in the presence of
  noise}, Mit Press Cambridge, Mass., 1963\relax
\mciteBstWouldAddEndPuncttrue
\mciteSetBstMidEndSepPunct{\mcitedefaultmidpunct}
{\mcitedefaultendpunct}{\mcitedefaultseppunct}\relax
\EndOfBibitem
\bibitem[Barlow and Levick(1965)]{barlow1965mechanism}
H.~Barlow and W.~R. Levick, \emph{The Journal of physiology}, 1965,
  \textbf{178}, 477--504\relax
\mciteBstWouldAddEndPuncttrue
\mciteSetBstMidEndSepPunct{\mcitedefaultmidpunct}
{\mcitedefaultendpunct}{\mcitedefaultseppunct}\relax
\EndOfBibitem
\bibitem[Gavrilov and Gavrilova(2001)]{gavrilov2001reliability}
L.~A. Gavrilov and N.~S. Gavrilova, \emph{Journal of theoretical Biology},
  2001, \textbf{213}, 527--545\relax
\mciteBstWouldAddEndPuncttrue
\mciteSetBstMidEndSepPunct{\mcitedefaultmidpunct}
{\mcitedefaultendpunct}{\mcitedefaultseppunct}\relax
\EndOfBibitem
\bibitem[Ferretti(2011)]{ferretti2011there}
P.~Ferretti, \emph{European Journal of Neuroscience}, 2011, \textbf{34},
  951--962\relax
\mciteBstWouldAddEndPuncttrue
\mciteSetBstMidEndSepPunct{\mcitedefaultmidpunct}
{\mcitedefaultendpunct}{\mcitedefaultseppunct}\relax
\EndOfBibitem
\bibitem[Moffet(2012)]{moffet2012nervous}
S.~B. Moffet, \emph{Nervous system regeneration in the invertebrates}, Springer
  Science \& Business Media, 2012, vol.~34\relax
\mciteBstWouldAddEndPuncttrue
\mciteSetBstMidEndSepPunct{\mcitedefaultmidpunct}
{\mcitedefaultendpunct}{\mcitedefaultseppunct}\relax
\EndOfBibitem
\bibitem[Levin(2007)]{levin2007large}
M.~Levin, \emph{Trends in cell biology}, 2007, \textbf{17}, 261--270\relax
\mciteBstWouldAddEndPuncttrue
\mciteSetBstMidEndSepPunct{\mcitedefaultmidpunct}
{\mcitedefaultendpunct}{\mcitedefaultseppunct}\relax
\EndOfBibitem
\bibitem[Levin(2009)]{levin2009bioelectric}
M.~Levin, Seminars in cell \& developmental biology, 2009, pp. 543--556\relax
\mciteBstWouldAddEndPuncttrue
\mciteSetBstMidEndSepPunct{\mcitedefaultmidpunct}
{\mcitedefaultendpunct}{\mcitedefaultseppunct}\relax
\EndOfBibitem
\bibitem[Tseng \emph{et~al.}(2010)Tseng, Beane, Lemire, Masi, and
  Levin]{tseng2010induction}
A.-S. Tseng, W.~S. Beane, J.~M. Lemire, A.~Masi and M.~Levin, \emph{Journal of
  Neuroscience}, 2010, \textbf{30}, 13192--13200\relax
\mciteBstWouldAddEndPuncttrue
\mciteSetBstMidEndSepPunct{\mcitedefaultmidpunct}
{\mcitedefaultendpunct}{\mcitedefaultseppunct}\relax
\EndOfBibitem
\bibitem[Eriksson \emph{et~al.}(1998)Eriksson, Perfilieva, Bj{\"o}rk-Eriksson,
  Alborn, Nordborg, Peterson, and Gage]{eriksson1998neurogenesis}
P.~S. Eriksson, E.~Perfilieva, T.~Bj{\"o}rk-Eriksson, A.-M. Alborn,
  C.~Nordborg, D.~A. Peterson and F.~H. Gage, \emph{Nature medicine}, 1998,
  \textbf{4}, 1313\relax
\mciteBstWouldAddEndPuncttrue
\mciteSetBstMidEndSepPunct{\mcitedefaultmidpunct}
{\mcitedefaultendpunct}{\mcitedefaultseppunct}\relax
\EndOfBibitem
\bibitem[van Praag \emph{et~al.}(2002)van Praag, Schinder, Christie, Toni,
  Palmer, and Gage]{van2002functional}
H.~van Praag, A.~F. Schinder, B.~R. Christie, N.~Toni, T.~D. Palmer and F.~H.
  Gage, \emph{Nature}, 2002, \textbf{415}, 1030\relax
\mciteBstWouldAddEndPuncttrue
\mciteSetBstMidEndSepPunct{\mcitedefaultmidpunct}
{\mcitedefaultendpunct}{\mcitedefaultseppunct}\relax
\EndOfBibitem
\bibitem[Hack \emph{et~al.}(2005)Hack, Saghatelyan, de~Chevigny, Pfeifer,
  Ashery-Padan, Lledo, and G{\"o}tz]{hack2005neuronal}
M.~A. Hack, A.~Saghatelyan, A.~de~Chevigny, A.~Pfeifer, R.~Ashery-Padan, P.-M.
  Lledo and M.~G{\"o}tz, \emph{Nature neuroscience}, 2005, \textbf{8},
  865\relax
\mciteBstWouldAddEndPuncttrue
\mciteSetBstMidEndSepPunct{\mcitedefaultmidpunct}
{\mcitedefaultendpunct}{\mcitedefaultseppunct}\relax
\EndOfBibitem
\bibitem[Zupanc(2006)]{zupanc2006neurogenesis}
G.~Zupanc, \emph{Journal of Comparative Physiology A}, 2006, \textbf{192},
  649\relax
\mciteBstWouldAddEndPuncttrue
\mciteSetBstMidEndSepPunct{\mcitedefaultmidpunct}
{\mcitedefaultendpunct}{\mcitedefaultseppunct}\relax
\EndOfBibitem
\bibitem[Tanaka and Ferretti(2009)]{tanaka2009considering}
E.~M. Tanaka and P.~Ferretti, \emph{Nature Reviews Neuroscience}, 2009,
  \textbf{10}, 713\relax
\mciteBstWouldAddEndPuncttrue
\mciteSetBstMidEndSepPunct{\mcitedefaultmidpunct}
{\mcitedefaultendpunct}{\mcitedefaultseppunct}\relax
\EndOfBibitem
\bibitem[Chen \emph{et~al.}(2006)Chen, Hall, and Chklovskii]{chen2006wiring}
B.~L. Chen, D.~H. Hall and D.~B. Chklovskii, \emph{Proceedings of the National
  Academy of Sciences}, 2006, \textbf{103}, 4723--4728\relax
\mciteBstWouldAddEndPuncttrue
\mciteSetBstMidEndSepPunct{\mcitedefaultmidpunct}
{\mcitedefaultendpunct}{\mcitedefaultseppunct}\relax
\EndOfBibitem
\bibitem[Chklovskii \emph{et~al.}(2002)Chklovskii, Schikorski, and
  Stevens]{chklovskii2002wiring}
D.~B. Chklovskii, T.~Schikorski and C.~F. Stevens, \emph{Neuron}, 2002,
  \textbf{34}, 341--347\relax
\mciteBstWouldAddEndPuncttrue
\mciteSetBstMidEndSepPunct{\mcitedefaultmidpunct}
{\mcitedefaultendpunct}{\mcitedefaultseppunct}\relax
\EndOfBibitem
\bibitem[Raj and Chen(2011)]{raj2011wiring}
A.~Raj and Y.-h. Chen, \emph{PloS one}, 2011, \textbf{6}, e14832\relax
\mciteBstWouldAddEndPuncttrue
\mciteSetBstMidEndSepPunct{\mcitedefaultmidpunct}
{\mcitedefaultendpunct}{\mcitedefaultseppunct}\relax
\EndOfBibitem
\bibitem[Miller \emph{et~al.}(1989)Miller, Todd, and
  Hegde]{miller1989designing}
G.~F. Miller, P.~M. Todd and S.~U. Hegde, ICGA, 1989, pp. 379--384\relax
\mciteBstWouldAddEndPuncttrue
\mciteSetBstMidEndSepPunct{\mcitedefaultmidpunct}
{\mcitedefaultendpunct}{\mcitedefaultseppunct}\relax
\EndOfBibitem
\bibitem[Angeline \emph{et~al.}(1994)Angeline, Saunders, and
  Pollack]{angeline1994evolutionary}
P.~J. Angeline, G.~M. Saunders and J.~B. Pollack, \emph{IEEE transactions on
  Neural Networks}, 1994, \textbf{5}, 54--65\relax
\mciteBstWouldAddEndPuncttrue
\mciteSetBstMidEndSepPunct{\mcitedefaultmidpunct}
{\mcitedefaultendpunct}{\mcitedefaultseppunct}\relax
\EndOfBibitem
\bibitem[Yao(1999)]{yao1999evolving}
X.~Yao, \emph{Proceedings of the IEEE}, 1999, \textbf{87}, 1423--1447\relax
\mciteBstWouldAddEndPuncttrue
\mciteSetBstMidEndSepPunct{\mcitedefaultmidpunct}
{\mcitedefaultendpunct}{\mcitedefaultseppunct}\relax
\EndOfBibitem
\bibitem[Floreano \emph{et~al.}(2008)Floreano, D{\"u}rr, and
  Mattiussi]{floreano2008neuroevolution}
D.~Floreano, P.~D{\"u}rr and C.~Mattiussi, \emph{Evolutionary Intelligence},
  2008, \textbf{1}, 47--62\relax
\mciteBstWouldAddEndPuncttrue
\mciteSetBstMidEndSepPunct{\mcitedefaultmidpunct}
{\mcitedefaultendpunct}{\mcitedefaultseppunct}\relax
\EndOfBibitem
\bibitem[Stanley and Miikkulainen(2002)]{stanley2002evolving}
K.~O. Stanley and R.~Miikkulainen, \emph{Evolutionary computation}, 2002,
  \textbf{10}, 99--127\relax
\mciteBstWouldAddEndPuncttrue
\mciteSetBstMidEndSepPunct{\mcitedefaultmidpunct}
{\mcitedefaultendpunct}{\mcitedefaultseppunct}\relax
\EndOfBibitem
\bibitem[Stanley \emph{et~al.}(2009)Stanley, D'Ambrosio, and
  Gauci]{stanley2009hypercube}
K.~O. Stanley, D.~B. D'Ambrosio and J.~Gauci, \emph{Artificial life}, 2009,
  \textbf{15}, 185--212\relax
\mciteBstWouldAddEndPuncttrue
\mciteSetBstMidEndSepPunct{\mcitedefaultmidpunct}
{\mcitedefaultendpunct}{\mcitedefaultseppunct}\relax
\EndOfBibitem
\bibitem[Eckmann \emph{et~al.}(2007)Eckmann, Feinerman, Gruendlinger, Moses,
  Soriano, and Tlusty]{eckmann2007physics}
J.-P. Eckmann, O.~Feinerman, L.~Gruendlinger, E.~Moses, J.~Soriano and
  T.~Tlusty, \emph{Physics Reports}, 2007, \textbf{449}, 54--76\relax
\mciteBstWouldAddEndPuncttrue
\mciteSetBstMidEndSepPunct{\mcitedefaultmidpunct}
{\mcitedefaultendpunct}{\mcitedefaultseppunct}\relax
\EndOfBibitem
\bibitem[Sz{\'e}kely(1965)]{szekely1965logical}
G.~Sz{\'e}kely, \emph{Acta physiologica Academiae Scientiarum Hungaricae},
  1965, \textbf{27}, 285\relax
\mciteBstWouldAddEndPuncttrue
\mciteSetBstMidEndSepPunct{\mcitedefaultmidpunct}
{\mcitedefaultendpunct}{\mcitedefaultseppunct}\relax
\EndOfBibitem
\bibitem[Stent \emph{et~al.}(1978)Stent, Kristan, Friesen, Ort, Poon, and
  Calabrese]{stent1978neuronal}
G.~S. Stent, W.~B. Kristan, W.~O. Friesen, C.~A. Ort, M.~Poon and R.~L.
  Calabrese, \emph{Science}, 1978, \textbf{200}, 1348--1357\relax
\mciteBstWouldAddEndPuncttrue
\mciteSetBstMidEndSepPunct{\mcitedefaultmidpunct}
{\mcitedefaultendpunct}{\mcitedefaultseppunct}\relax
\EndOfBibitem
\bibitem[Friesen and Stent(1978)]{friesen1978neural}
W.~O. Friesen and G.~S. Stent, \emph{Annual review of biophysics and
  bioengineering}, 1978, \textbf{7}, 37--61\relax
\mciteBstWouldAddEndPuncttrue
\mciteSetBstMidEndSepPunct{\mcitedefaultmidpunct}
{\mcitedefaultendpunct}{\mcitedefaultseppunct}\relax
\EndOfBibitem
\bibitem[Amari(1977)]{amari1977dynamics}
S.-i. Amari, \emph{Biological cybernetics}, 1977, \textbf{27}, 77--87\relax
\mciteBstWouldAddEndPuncttrue
\mciteSetBstMidEndSepPunct{\mcitedefaultmidpunct}
{\mcitedefaultendpunct}{\mcitedefaultseppunct}\relax
\EndOfBibitem
\bibitem[Clune \emph{et~al.}(2013)Clune, Mouret, and
  Lipson]{clune2013evolutionary}
J.~Clune, J.-B. Mouret and H.~Lipson, \emph{Proceedings of the Royal Society b:
  Biological sciences}, 2013, \textbf{280}, 20122863\relax
\mciteBstWouldAddEndPuncttrue
\mciteSetBstMidEndSepPunct{\mcitedefaultmidpunct}
{\mcitedefaultendpunct}{\mcitedefaultseppunct}\relax
\EndOfBibitem
\bibitem[Floreano \emph{et~al.}(2001)Floreano, Nolfi, and
  Mondada]{floreano2001co}
D.~Floreano, S.~Nolfi and F.~Mondada, \emph{Advances in the evolutionary
  synthesis of intelligent agents}, 2001,  273--306\relax
\mciteBstWouldAddEndPuncttrue
\mciteSetBstMidEndSepPunct{\mcitedefaultmidpunct}
{\mcitedefaultendpunct}{\mcitedefaultseppunct}\relax
\EndOfBibitem
\bibitem[Pfeifer \emph{et~al.}(2007)Pfeifer, Lungarella, and
  Iida]{pfeifer2007self}
R.~Pfeifer, M.~Lungarella and F.~Iida, \emph{science}, 2007, \textbf{318},
  1088--1093\relax
\mciteBstWouldAddEndPuncttrue
\mciteSetBstMidEndSepPunct{\mcitedefaultmidpunct}
{\mcitedefaultendpunct}{\mcitedefaultseppunct}\relax
\EndOfBibitem
\bibitem[Coello \emph{et~al.}(2006)Coello, De~Computaci{\'o}n, and
  Zacatenco]{coello2006twenty}
C.~Coello, S.~De~Computaci{\'o}n and C.~Zacatenco, \emph{IEEE computational
  intelligence magazine}, 2006, \textbf{1}, 28--36\relax
\mciteBstWouldAddEndPuncttrue
\mciteSetBstMidEndSepPunct{\mcitedefaultmidpunct}
{\mcitedefaultendpunct}{\mcitedefaultseppunct}\relax
\EndOfBibitem
\bibitem[Schuster(2012)]{schuster2012optimization}
P.~Schuster, \emph{Complexity}, 2012, \textbf{18}, 5--7\relax
\mciteBstWouldAddEndPuncttrue
\mciteSetBstMidEndSepPunct{\mcitedefaultmidpunct}
{\mcitedefaultendpunct}{\mcitedefaultseppunct}\relax
\EndOfBibitem
\bibitem[Seoane
  \emph{et~al.}(2016)Seoane\emph{et~al.}]{seoane2016multiobjetive}
L.~F. Seoane \emph{et~al.}, \emph{Ph.D. thesis}, Universitat Pompeu Fabra,
  2016\relax
\mciteBstWouldAddEndPuncttrue
\mciteSetBstMidEndSepPunct{\mcitedefaultmidpunct}
{\mcitedefaultendpunct}{\mcitedefaultseppunct}\relax
\EndOfBibitem
\bibitem[Seoane and Sol{\'e}(2013)]{seoane2013multiobjective}
L.~F. Seoane and R.~V. Sol{\'e}, \emph{arXiv preprint arXiv:1310.6372},
  2013\relax
\mciteBstWouldAddEndPuncttrue
\mciteSetBstMidEndSepPunct{\mcitedefaultmidpunct}
{\mcitedefaultendpunct}{\mcitedefaultseppunct}\relax
\EndOfBibitem
\bibitem[Seoane and Sol{\'e}(2015)]{seoane2015phase}
L.~F. Seoane and R.~Sol{\'e}, \emph{Physical Review E}, 2015, \textbf{92},
  032807\relax
\mciteBstWouldAddEndPuncttrue
\mciteSetBstMidEndSepPunct{\mcitedefaultmidpunct}
{\mcitedefaultendpunct}{\mcitedefaultseppunct}\relax
\EndOfBibitem
\bibitem[Seoane and Sol{\'e}(2015)]{seoane2015systems}
L.~F. Seoane and R.~Sol{\'e}, \emph{arXiv preprint arXiv:1510.08697},
  2015\relax
\mciteBstWouldAddEndPuncttrue
\mciteSetBstMidEndSepPunct{\mcitedefaultmidpunct}
{\mcitedefaultendpunct}{\mcitedefaultseppunct}\relax
\EndOfBibitem
\bibitem[Seoane and Sol{\'e}(2016)]{seoane2016multiobjective}
L.~F. Seoane and R.~Sol{\'e}, in \emph{Proceedings of ECCS 2014}, Springer,
  2016, pp. 259--270\relax
\mciteBstWouldAddEndPuncttrue
\mciteSetBstMidEndSepPunct{\mcitedefaultmidpunct}
{\mcitedefaultendpunct}{\mcitedefaultseppunct}\relax
\EndOfBibitem
\bibitem[Seoane and Sol{\'e}(2018)]{seoane2018morphospace}
L.~F. Seoane and R.~Sol{\'e}, \emph{Scientific reports}, 2018, \textbf{8},
  year\relax
\mciteBstWouldAddEndPuncttrue
\mciteSetBstMidEndSepPunct{\mcitedefaultmidpunct}
{\mcitedefaultendpunct}{\mcitedefaultseppunct}\relax
\EndOfBibitem
\bibitem[Rojas(2013)]{rojas2013neural}
R.~Rojas, \emph{Neural networks: a systematic introduction}, Springer Science
  \& Business Media, 2013\relax
\mciteBstWouldAddEndPuncttrue
\mciteSetBstMidEndSepPunct{\mcitedefaultmidpunct}
{\mcitedefaultendpunct}{\mcitedefaultseppunct}\relax
\EndOfBibitem
\bibitem[Konak \emph{et~al.}(2006)Konak, Coit, and Smith]{konak2006multi}
A.~Konak, D.~W. Coit and A.~E. Smith, \emph{Reliability Engineering \& System
  Safety}, 2006, \textbf{91}, 992--1007\relax
\mciteBstWouldAddEndPuncttrue
\mciteSetBstMidEndSepPunct{\mcitedefaultmidpunct}
{\mcitedefaultendpunct}{\mcitedefaultseppunct}\relax
\EndOfBibitem
\bibitem[Tononi \emph{et~al.}(1999)Tononi, Sporns, and
  Edelman]{tononi1999measures}
G.~Tononi, O.~Sporns and G.~M. Edelman, \emph{Proceedings of the National
  Academy of Sciences}, 1999, \textbf{96}, 3257--3262\relax
\mciteBstWouldAddEndPuncttrue
\mciteSetBstMidEndSepPunct{\mcitedefaultmidpunct}
{\mcitedefaultendpunct}{\mcitedefaultseppunct}\relax
\EndOfBibitem
\bibitem[Edelman and Gally(2001)]{edelman2001degeneracy}
G.~M. Edelman and J.~A. Gally, \emph{Proceedings of the National Academy of
  Sciences}, 2001, \textbf{98}, 13763--13768\relax
\mciteBstWouldAddEndPuncttrue
\mciteSetBstMidEndSepPunct{\mcitedefaultmidpunct}
{\mcitedefaultendpunct}{\mcitedefaultseppunct}\relax
\EndOfBibitem
\bibitem[Mac{\'\i}a \emph{et~al.}(2012)Mac{\'\i}a, Posas, and
  Sol{\'e}]{macia2012distributed}
J.~Mac{\'\i}a, F.~Posas and R.~V. Sol{\'e}, \emph{Trends in biotechnology},
  2012, \textbf{30}, 342--349\relax
\mciteBstWouldAddEndPuncttrue
\mciteSetBstMidEndSepPunct{\mcitedefaultmidpunct}
{\mcitedefaultendpunct}{\mcitedefaultseppunct}\relax
\EndOfBibitem
\bibitem[Regot \emph{et~al.}(2011)Regot, Macia, Conde, Furukawa, Kjell{\'e}n,
  Peeters, Hohmann, De~Nadal, Posas, and Sol{\'e}]{regot2011distributed}
S.~Regot, J.~Macia, N.~Conde, K.~Furukawa, J.~Kjell{\'e}n, T.~Peeters,
  S.~Hohmann, E.~De~Nadal, F.~Posas and R.~Sol{\'e}, \emph{Nature}, 2011,
  \textbf{469}, 207\relax
\mciteBstWouldAddEndPuncttrue
\mciteSetBstMidEndSepPunct{\mcitedefaultmidpunct}
{\mcitedefaultendpunct}{\mcitedefaultseppunct}\relax
\EndOfBibitem
\bibitem[Jaeger(2001)]{jaeger2001echo}
H.~Jaeger, \emph{Bonn, Germany: German National Research Center for Information
  Technology GMD Technical Report}, 2001, \textbf{148}, 13\relax
\mciteBstWouldAddEndPuncttrue
\mciteSetBstMidEndSepPunct{\mcitedefaultmidpunct}
{\mcitedefaultendpunct}{\mcitedefaultseppunct}\relax
\EndOfBibitem
\bibitem[Maass \emph{et~al.}(2002)Maass, Natschl{\"a}ger, and
  Markram]{maass2002real}
W.~Maass, T.~Natschl{\"a}ger and H.~Markram, \emph{Neural computation}, 2002,
  \textbf{14}, 2531--2560\relax
\mciteBstWouldAddEndPuncttrue
\mciteSetBstMidEndSepPunct{\mcitedefaultmidpunct}
{\mcitedefaultendpunct}{\mcitedefaultseppunct}\relax
\EndOfBibitem
\bibitem[Jaeger \emph{et~al.}(2007)Jaeger, Maass, and
  Principe]{jaeger2007special}
H.~Jaeger, W.~Maass and J.~Principe, 2007\relax
\mciteBstWouldAddEndPuncttrue
\mciteSetBstMidEndSepPunct{\mcitedefaultmidpunct}
{\mcitedefaultendpunct}{\mcitedefaultseppunct}\relax
\EndOfBibitem
\bibitem[Verstraeten \emph{et~al.}(2007)Verstraeten, Schrauwen, d’Haene, and
  Stroobandt]{verstraeten2007experimental}
D.~Verstraeten, B.~Schrauwen, M.~d’Haene and D.~Stroobandt, \emph{Neural
  networks}, 2007, \textbf{20}, 391--403\relax
\mciteBstWouldAddEndPuncttrue
\mciteSetBstMidEndSepPunct{\mcitedefaultmidpunct}
{\mcitedefaultendpunct}{\mcitedefaultseppunct}\relax
\EndOfBibitem
\bibitem[Luko{\v{s}}evi{\v{c}}ius \emph{et~al.}(2012)Luko{\v{s}}evi{\v{c}}ius,
  Jaeger, and Schrauwen]{lukovsevivcius2012reservoir}
M.~Luko{\v{s}}evi{\v{c}}ius, H.~Jaeger and B.~Schrauwen,
  \emph{KI-K{\"u}nstliche Intelligenz}, 2012, \textbf{26}, 365--371\relax
\mciteBstWouldAddEndPuncttrue
\mciteSetBstMidEndSepPunct{\mcitedefaultmidpunct}
{\mcitedefaultendpunct}{\mcitedefaultseppunct}\relax
\EndOfBibitem
\bibitem[Sol{\'e}(2011)]{sole2011phase}
R.~Sol{\'e}, \emph{Press. Princeton}, 2011\relax
\mciteBstWouldAddEndPuncttrue
\mciteSetBstMidEndSepPunct{\mcitedefaultmidpunct}
{\mcitedefaultendpunct}{\mcitedefaultseppunct}\relax
\EndOfBibitem
\bibitem[Oll\'e-Vila \emph{et~al.}(2016)Oll\'e-Vila, Duran-Nebreda,
  Conde-Pueyo, Mont\'a\~{n}ez, and Sol\'e]{olle2016morphospace}
A.~Oll\'e-Vila, S.~Duran-Nebreda, N.~Conde-Pueyo, R.~Mont\'a\~{n}ez and
  R.~Sol\'e, \emph{Integrative Biology}, 2016, \textbf{8}, 485--503\relax
\mciteBstWouldAddEndPuncttrue
\mciteSetBstMidEndSepPunct{\mcitedefaultmidpunct}
{\mcitedefaultendpunct}{\mcitedefaultseppunct}\relax
\EndOfBibitem
\end{mcitethebibliography}

\end{document}